\documentstyle[prb,aps,preprint]{revtex}
\begin{document}
\draft
\preprint{YITP--96--14}
\title{Tunneling through a barrier in Tomonaga-Luttinger 
liquid connected to reservoirs }
\author{Akira Furusaki}
\address{Yukawa Institute for Theoretical Physics, Kyoto University,
Kyoto 606-01, Japan}
\author{Naoto Nagaosa}
\address{Department of Applied Physics, University of Tokyo,
Bunkyo-ku, Tokyo 113, Japan}
\date{April 24, 1996}
\maketitle
\begin{abstract}
The transport through a barrier is studied for a Tomonaga-Luttinger liquid 
of finite length connected to reservoirs.
An effective action for the phase variable at the barrier is derived
for spatially varying electric field.
In the d.c.\ limit only the total voltage drop 
between the left and right reservoirs appears in the action.
We discuss crossovers of the renormalization group flow and hence of
the temperature and wire length dependence of the conductance,
taking into account the location of the barrier.
\end{abstract}

\narrowtext
Recently the quantum transport in Tomonaga-Luttinger (TL) liquids has
been studied intensively both experimentally and theoretically
\cite{kf,fn,mat,nag1,of}.
One of the subtle issues in this problem is the effects of the leads and 
contacts which are often ignored in the treatment of infinitely long TL wires.
Several authors reached the conclusion that the renormalization of the 
conductance $G$ due to the electron-electron interaction is absent in the
clean TL liquids when the reservoirs, i.e., leads, are properly taken into
account \cite{mas1,mas2,pon,safi,note}.
The same conclusion has been obtained even for the infinite TL wire by 
considering carefully the definition of the voltage drop, i.e., the chemical
potential difference between the right and left leads \cite{kawa,si}.
Inspired by these recent works, we examine in this paper the transport 
through a barrier in a finite-length quantum wire connected to the leads.
We find nontrivial dependence of the conductance on the temperature,
wire length, and also the location of the barrier.

Our model is similar to those of Ref.~\cite{mas1,mas2}, and
its Lagrangian in the imaginary time is given as
\begin{eqnarray}
L &=& \int d x { 1 \over {2 K(x)} } \biggl[
 { 1 \over {v(x)}} \biggl( { { \partial \phi} \over {\partial \tau} } 
 \biggr)^2 + 
 v(x) \biggl( { { \partial \phi} \over {\partial x} } \biggr)^2
 \biggr] 
\nonumber \\
&& + { \lambda_B \over { \pi \alpha} } 
 \cos [ 2 k_F a + 2 \sqrt{\pi} \phi(a,\tau)]
+ \int d x E(x,\tau) { 1 \over {\sqrt{\pi}} } \phi(x,\tau),
\end{eqnarray}
where $K(x)$ and $v(x)$ is the spatially varying exponent and
velocity of the TL liquid, and $\alpha$ is a short-distance cutoff.
Following Ref.~\cite{mas1}, we assume that the wire is confined in $0<x<L$
and the leads extend for $x<0$ and $x>L$.
Correspondingly $K(x) = K_W$, $v(x) = v_W$ for the wire and 
$K(x) = K_L = 1$, $v(x) = v_L$ for the leads.
The second term on the rhs of Eq.~(1) describes the backward scattering by
the barrier potential at $x=a$ ($0<a<L$) whose strength is $\lambda_B$.
The third term is the coupling with the electric field $E(x)$.
Now we integrate over the continuum degrees of freedom $\phi(x,\tau)$
in the functional integral with the fixed value of 
$\phi(x=a,\tau)= \phi_0(\tau)$ and $E(x, \tau)$.
This procedure can be done if one knows the Green's function
$G_{\omega_n}(x,x')$ for the unperturbed TL liquid described by the first
term in Eq.~(1).
Note that it is no longer a function of $x-x'$ because the translational
symmetry is broken in our system.
In fact $G_{\omega_n}(x,x')$ has been obtained in Ref.~\cite{mas1}, but here
we derive the effective action first without refering to its explicit form:
\begin{eqnarray}
S &=& { 1 \over {2 \beta} } \sum_{\omega_n}
{1 \over {G_{\omega_n} (a,a) }} \tilde\phi_0(-\omega_n) \tilde\phi_0(\omega_n)
 + 
{ \lambda_B \over { \pi \alpha} } \int^\beta_0d\tau
 \cos [ 2 k_F a + 2 \sqrt{\pi} \phi_0(\tau)]
\nonumber \\
& & + \sum_{\omega_n} \int dx  
{ { G_{\omega_n}(x,a) } \over  
{  \sqrt{\pi} G_{\omega_n}(a,a) } }
\tilde E(x, \omega_n) \tilde\phi_0(-\omega_n)
\nonumber \\
& & - {\beta \over { 2 \pi}}
\sum_{\omega_n} \int dx \int d x' 
\biggl[ 
G_{\omega_n}(x,x') - 
{ { G_{\omega_n}(x,a) G_{\omega_n}(x',a)} \over  
{G_{\omega_n}(a,a) } } \biggr]
\tilde E(x,\omega_n) \tilde E(x',-\omega_n),
\end{eqnarray}
where $\tilde X(\omega_n)=\int d\tau e^{i\omega_n\tau}X(\tau)$
($X=\phi_0$ and $E$).
In this calculation we used the relation
$G_\omega(x,x')=G_{|\omega|}(x,x')=G_{|\omega|}(x',x)$.
Only the following information is needed for our purpose:
\begin{equation}
\lim_{\omega \to 0} G_\omega ( x,x') = { {K_L} \over { 2 |\omega|} },
\end{equation}
\begin{equation}
G_{\omega} ( a,a) = { K_W \over { 2 |\omega|} }
+ { K_W \over {|\omega|} } { 
{ (K_L-K_W)^2 e^{-L/L_\omega} + (K_L^2-K_W^2)
\cosh[(L-2a)/L_\omega] }
\over 
{ (K_L + K_W)^2 e^{L/L_\omega} - (K_L-K_W)^2 e^{-L/L_\omega} }},
\end{equation}
where $L_{\omega} = v_W/|\omega|$.

Here we are interested in the d.c.\ conductance, i.e., 
$\tilde E(x,\omega_n) = E(x) \delta_{\omega_n,0}$ and 
from Eq.~(3) the action becomes
\begin{eqnarray}
S &=& { 1 \over {2 \beta} }  \sum_{\omega_n} 
{ 1 \over {G_{\omega_n} (a,a) }} 
\tilde\phi_0(-\omega_n) \tilde\phi_0(\omega_n)
 + 
{ \lambda_B \over { \pi \alpha} } \int^\beta_0d\tau
 \cos [ 2 k_F a + 2 \sqrt{\pi} \phi_0(\tau)]
\nonumber \\
&& + { 1 \over  {\sqrt{\pi}} }  \Delta V \tilde\phi_0(\omega_n=0).
\end{eqnarray}
Note that only the total voltage drop 
$\Delta V = \int_{-\infty}^{\infty} dx E(x)$ appears in the action,
which couples linearly with the phase variable at the barrier. 
Thus we can calculate the d.c.~conductance without the detailed knowledge
of the spatial dependence of the electric field $E(x)$, which should be
determined by solving the Maxwell equations self-consistently.

We study the action Eq.~(5) in the limit of weak and strong barrier
potential by employing the renormalization group (RG) method.
Since the analysis is now standard, we skip the detailed derivation and
give only the final results and their physical picture.
Assuming $0<a<L/2$ without loss of generality, we can approximate Eq.~(4)
as
\begin{equation}
G_{\omega} (a,a) \cong { K \over { 2 |\omega|} },
\end{equation}
where the effective exponent $K$ is given by
\begin{equation}
K = \cases{
 K_W & $|\omega_n|\gg v_W/a$, \cr
 { 2 K_L K_W \over { K_W + K_L} } & $v_W/L\ll|\omega_n|\ll v_W/a$, \cr
 K_L & $|\omega_n|\ll v_W/L$, \cr}
\end{equation}
where we have assumed $a\ll L$.
When $a\sim L/2$ the second region $v_W/L\ll|\omega_n|\ll v_W/a$ collapses.
As the frequency $|\omega_n|$ decreases, the longer range properties
become relevant.
In the low-frequency limit the presence of the Fermi-liquid leads is
essential, while in the high-frequency regime the exponent of the TL wire
controls the renormalization.
In the intermediate frequency regime, both $K_W$ and $K_L$ contribute
because $\phi_0$ {\it sees} both the Fermi-liquid lead ($x<0$) and
the TL wire ($a<x<L$).

First we consider the weak-potential limit.
The RG equation for $\lambda_B$ reads
\begin{equation}
{ {d \lambda_B} \over {d l}} = ( 1 -K) \lambda_B,
\end{equation}
where $l = - \ln \Lambda$ ($\Lambda$: frequency cutoff ).
Up to second order in $\lambda_B$ the conductance $G(T)$ at temperature $T$
is then given by
\begin{equation}
G(T) = \cases{
{ e^2 \over { 2 \pi}} - c_1 e^2 
\left( { \lambda_B \over v_W } \right)^2
\left( { T \over {D} } \right)^{2 K_W -2}
& $T\gg v_W/a$, \cr
{ e^2 \over { 2 \pi}} - c_2 e^2
\left( { \lambda_B \over v_W } \right)^2
\left( { v_W \over aD } \right)^{2 K_W -2}
\left( { aT  \over v_W } \right)^{2(K_W -1)/(K_W + 1)}
& $v_W/L\ll T\ll v_W/a$, \cr
{ e^2 \over { 2 \pi}} - c_3 e^2
\left( { \lambda_B \over v_W } \right)^2
\left( { v_W \over aD } \right)^{2 K_W -2}
\left( { a \over L } \right)^{2(K_W -1)/(K_W + 1)}
& $T\ll v_W/L$, \cr}
\end{equation}
where $c$'s are constants of order unity and $D$ is a high-energy cutoff
(band width, $\alpha D\sim v_W$).
As we see in Eq.~(9), in the presence of a single barrier near the edge
of the wire ($\alpha\ll a\ll L$), the conductance shows the power-law
temperature dependence characteristic of the TL liquid in the high
and intermediate temperature regime.
In the low-temperature limit, the conductance does not depend on $T$
because the renormalization is cut off by the finite length of the wire.
If the barrier is in the middle of the wire ($a\approx L/2$), the
intermediate temperature regime does not exist, and furthermore
in the low-temperature regime the correction term proportional to
$\lambda^2_B$ depends on the wire length as $L^{2-2K_W}$.
When {\it many weak} impurities (barriers) are distributed in the wire,
we should average over the location $a$ of the barriers.
For a unifrom distribution, there remains only a single crossover
temperature $v_W/L$
because the dominant contribution comes from the case $a \sim L/2$.
We thus reproduce the result of Maslov, Eq.~(14) in Ref.~\cite{mas2}.
We note that Eq.~(9) is valid only when the second term on the rhs
is much smaller than the first term, $e^2/2\pi$.
This condition is the most severe at low temperatures, and 
is always satisfied if
$\left( { \lambda_B \over v_W } \right)^2
\left( { v_W \over aD } \right)^{2 K_W -2}
\left( { a \over L } \right)^{(K_W -1)/(K_W + 1)}\ll 1$.
Otherwise we can use Eq.~(9) only above some temperature, below which
the barrier potential should be regarded as strong.

We now turn to the limit of strong barrier potential, which is not
discussed in Ref.~\cite{mas2}.
The condition for this limit to be realized is
$D(\lambda_B/v_W)^{1/(1-K_W)}\gg v_W/a$.
Duality mapping using the instantons \cite{smd} is useful in this case,
and the effective acton for the dual field $\theta_0(\tau)$ is given by  
\begin{equation}
S = { 2 \over \beta} \sum_{\omega_n}
 \omega_n^2 G_{\omega_n}(a,a)
 \tilde\theta_0(-\omega_n) \tilde\theta_0(\omega_n)
 + 2t \int_0^{\beta} d \tau \cos [ 2 \sqrt{\pi} \theta_0(\tau)],
\end{equation}
where $t$ is the fugacity of the instanton, i.e., the tunneling matrix
element through the barrier.
The RG equation for $t$ is readily obtained as
\begin{equation}
{ {d t} \over {d l}} = \left( 1 - {1 \over K} \right) t,
\end{equation}
where $K$ is given in Eq.~(7).
Then it is straightforward to derive the temperture
dependence of the conductance $G(T)$ in this limit:
\begin{equation}
G(T) = \cases{
{\tilde c}_1 e^2 
\left( { t \over D } \right)^2
\left( { T \over D } \right)^{2/K_W - 2}
& $T\gg v_W/a$, \cr
{\tilde c}_2 e^2 
\left( { t \over D } \right)^2
\left( { {v_W T}  \over {a D^2} } \right)^{1/K_W -1}
& $v_W/L\ll T\ll v_W/a$, \cr
{\tilde c}_3 e^2 
\left( { t \over D } \right)^2
\left( { {v_W^2 }  \over {aL D^2} } \right)^{1/K_W -1}
& $T\ll v_W/L$. \cr}
\end{equation}
Equation (12) is valid for $\alpha\ll a\ll L/2$.
If $a\approx L/2$, the intermediate temperature regime does not exist,
and $G(T)\propto L^{2-2/K_W}$ in the low-temperature limit.
Since the strongest barrier determines the transport of the whole system,
at low temperatures where the impurity potentials become strong due to
the renormalization, the $T$- and/or $L$-dependence of the conductance
shown in Eq.~(12) might be observed in experiments of long
quantum wires.

So far we have discussed the transport of spinless TL liquids.
To compare the theory with experiments in quantum wires, we only
need to replace the RG equations, Eqs.~(8) and (11), by
$\frac{d\lambda_B}{dl}=\frac12(1-K)\lambda_B$ and
$\frac{dt}{dl}=\frac12(1-\frac1K)t$.
Accordingly, the exponents of the renormalization factors in Eqs.~(9) and
(12) should be divided by 2.

Finally we briefly comment on the effect of the Coulomb interaction.
With this interaction the effective exponent $K$ is a function of the energy
even for the infinite wire and goes to zero in the low-energy limit.
For the finite-length wires the RG is cut off by the energy scale
$v_W/L$ or $v_W/a$.
For example, when a single strong barrier is present in the middle of
the wire, the conductance should show the anomalous temperature dependence
discussed in Ref.~\cite{fn} at $T\gg v_W/L$,
$G(T)\propto T^{-2}\exp\!\left[-A\left(\ln v_W/WT\right)^{3/2}\right]$
($A$ is a constant and $W$ is the width of the wire).
At lower temperatures $T\ll v_W/L$, $T$ should be replaced by $v_W/L$ and
the conductance becomes independent of $T$.
We note, however, that the above discussion is possibly oversimplified,
and the screening of the Coulomb interaction due to the electrons in
the Fermi-liquid leads and the gates near the wire should be properly
taken into account.

In summary we have studied the tunneling through a barrier in a
finite-length TL liquid connceted to Fermi-liquid leads. 
The conductance shows a peculiar dependence on the temperature, the wire
length, and the location of the barrier.
We propose that systematic studies of these dependence will give a firm
evidence for the TL liquid in the quantum wire.

\acknowledgements
The authors acknowledge A.\ M.\ Finkel'stein, H.\ Fukuyama,
M.\ Ogata, and V.\ V.\ Ponomarenko for useful discussions.
This work was supported by Grant-in-Aid for 
Scientific Research (No.~06243103)
from the Ministry of Education, Science, and Culture of Japan.

\end{document}